\documentclass[12pt,preprint]{aastex}

\slugcomment{Accepted by ApJ for publication}

\begin{document}

\title{Identification of the TeV Gamma-ray Source  ARGO J2031+4157 with the Cygnus Cocoon}



\author{B.~Bartoli\altaffilmark{1,2},
 P.~Bernardini\altaffilmark{3,4},
 X.J.~Bi\altaffilmark{5},
 P.~Branchini\altaffilmark{6},
 A.~Budano\altaffilmark{6},
 P.~Camarri\altaffilmark{7,8},
 Z.~Cao\altaffilmark{5},
 R.~Cardarelli\altaffilmark{8},
 S.~Catalanotti\altaffilmark{1,2},
 S.Z.~Chen\altaffilmark{0,5},\footnotetext[0]{Corresponding author: Songzhan Chen, chensz@ihep.ac.cn},
 T.L.~Chen\altaffilmark{9},
 P.~Creti\altaffilmark{4},
 S.W.~Cui\altaffilmark{10},
 B.Z.~Dai\altaffilmark{11},
 A.~D'Amone\altaffilmark{3,4},
 Danzengluobu\altaffilmark{9},
 I.~De Mitri\altaffilmark{3,4},
 B.~D'Ettorre Piazzoli\altaffilmark{1,2},
 T.~Di Girolamo\altaffilmark{1,2},
 G.~Di Sciascio\altaffilmark{8},
 C.F.~Feng\altaffilmark{12},
 Zhaoyang Feng\altaffilmark{5},
 Zhenyong Feng\altaffilmark{13},
 Q.B.~Gou\altaffilmark{5},
 Y.Q.~Guo\altaffilmark{5},
 H.H.~He\altaffilmark{5},
 Haibing Hu\altaffilmark{9},
 Hongbo Hu\altaffilmark{5},
 M.~Iacovacci\altaffilmark{1,2},
 R.~Iuppa\altaffilmark{7,8},
 H.Y.~Jia\altaffilmark{13},
 Labaciren\altaffilmark{9},
 H.J.~Li\altaffilmark{9},
 G.~Liguori\altaffilmark{14,15},
 C.~Liu\altaffilmark{5},
 J.~Liu\altaffilmark{11},
 M.Y.~Liu\altaffilmark{9},
 H.~Lu\altaffilmark{5},
 L.L.~Ma\altaffilmark{5},
 X.H.~Ma\altaffilmark{5},
 G.~Mancarella\altaffilmark{3,4},
 S.M.~Mari\altaffilmark{6,16},
 G.~Marsella\altaffilmark{3,4},
 D.~Martello\altaffilmark{3,4},
 S.~Mastroianni\altaffilmark{2},
 P.~Montini\altaffilmark{6,16},
 C.C.~Ning\altaffilmark{9},
 M.~Panareo\altaffilmark{3,4},
 L.~Perrone\altaffilmark{3,4},
 P.~Pistilli\altaffilmark{6,16},
 F.~Ruggieri\altaffilmark{6},
 P.~Salvini\altaffilmark{15},
 R.~Santonico\altaffilmark{7,8},
 P.R.~Shen\altaffilmark{5},
 X.D.~Sheng\altaffilmark{5},
 F.~Shi\altaffilmark{5},
 A.~Surdo\altaffilmark{4},
 Y.H.~Tan\altaffilmark{5},
 P.~Vallania\altaffilmark{17,18},
 S.~Vernetto\altaffilmark{17,18},
 C.~Vigorito\altaffilmark{18,19},
 H.~Wang\altaffilmark{5},
 C.Y.~Wu\altaffilmark{5},
 H.R.~Wu\altaffilmark{5},
 L.~Xue\altaffilmark{12},
 Q.Y.~Yang\altaffilmark{11},
 X.C.~Yang\altaffilmark{11},
 Z.G.~Yao\altaffilmark{5},
 A.F.~Yuan\altaffilmark{9},
 M.~Zha\altaffilmark{5},
 H.M.~Zhang\altaffilmark{5},
 L.~Zhang\altaffilmark{11},
 X.Y.~Zhang\altaffilmark{12},
 Y.~Zhang\altaffilmark{5},
 J.~Zhao\altaffilmark{5},
 Zhaxiciren\altaffilmark{9},
 Zhaxisangzhu\altaffilmark{9},
 X.X.~Zhou\altaffilmark{13},
 F.R.~Zhu\altaffilmark{13},
 Q.Q.~Zhu\altaffilmark{5} and
 G.~Zizzi\altaffilmark{20}\\ (The ARGO-YBJ Collaboration)}


 \altaffiltext{1}{Dipartimento di Fisica dell'Universit\`a di Napoli
                  ``Federico II'', Complesso Universitario di Monte
                  Sant'Angelo, via Cinthia, 80126 Napoli, Italy.}
 \altaffiltext{2}{Istituto Nazionale di Fisica Nucleare, Sezione di
                  Napoli, Complesso Universitario di Monte
                  Sant'Angelo, via Cinthia, 80126 Napoli, Italy.}
 \altaffiltext{3}{Dipartimento Matematica e Fisica "Ennio De Giorgi",
                  Universit\`a del Salento,
                  via per Arnesano, 73100 Lecce, Italy.}
 \altaffiltext{4}{Istituto Nazionale di Fisica Nucleare, Sezione di
                  Lecce, via per Arnesano, 73100 Lecce, Italy.}
 \altaffiltext{5}{Key Laboratory of Particle Astrophysics, Institute
                  of High Energy Physics, Chinese Academy of Sciences,
                  P.O. Box 918, 100049 Beijing, P.R. China.}
 \altaffiltext{6}{Istituto Nazionale di Fisica Nucleare, Sezione di
                  Roma Tre, via della Vasca Navale 84, 00146 Roma, Italy.}
 \altaffiltext{7}{Dipartimento di Fisica dell'Universit\`a di Roma
                  ``Tor Vergata'', via della Ricerca Scientifica 1,
                  00133 Roma, Italy.}
 \altaffiltext{8}{Istituto Nazionale di Fisica Nucleare, Sezione di
                  Roma Tor Vergata, via della Ricerca Scientifica 1,
                  00133 Roma, Italy.}
 \altaffiltext{9}{Tibet University, 850000 Lhasa, Xizang, P.R. China.}
 \altaffiltext{10}{Hebei Normal University, Shijiazhuang 050016,
                   Hebei, P.R. China.}
 \altaffiltext{11}{Yunnan University, 2 North Cuihu Rd., 650091 Kunming,
                   Yunnan, P.R. China.}
 \altaffiltext{12}{Shandong University, 250100 Jinan, Shandong, P.R. China.}
 \altaffiltext{13}{Southwest Jiaotong University, 610031 Chengdu,
                   Sichuan, P.R. China.}
 \altaffiltext{14}{Dipartimento di Fisica dell'Universit\`a di
                   Pavia, via Bassi 6, 27100 Pavia, Italy.}
 \altaffiltext{15}{Istituto Nazionale di Fisica Nucleare, Sezione di Pavia,
                   via Bassi 6, 27100 Pavia, Italy.}
 \altaffiltext{16}{Dipartimento di Fisica dell'Universit\`a ``Roma Tre'',
                   via della Vasca Navale 84, 00146 Roma, Italy.}
 \altaffiltext{17}{Osservatorio Astrofisico di Torino dell'Istituto Nazionale
                   di Astrofisica, via P. Giuria 1, 10125 Torino, Italy.}
 \altaffiltext{18}{Istituto Nazionale di Fisica Nucleare,
                   Sezione di Torino, via P. Giuria 1, 10125 Torino, Italy.}
 \altaffiltext{19}{Dipartimento di Fisica dell'Universit\`a di
                   Torino, via P. Giuria 1, 10125 Torino, Italy.}
 \altaffiltext{20}{Istituto Nazionale di Fisica Nucleare - CNAF, Viale
                   Berti-Pichat 6/2, 40127 Bologna, Italy.}

\begin{abstract}
The extended TeV gamma-ray source ARGO J2031+4157 (or MGRO J2031+41) is positionally consistent with the Cygnus Cocoon discovered by $Fermi$-LAT at GeV energies in the Cygnus superbubble.
Reanalyzing the ARGO-YBJ data collected from November 2007 to January 2013,
the angular extension and energy spectrum of ARGO J2031+4157 are evaluated.
After subtracting the contribution of the overlapping TeV sources,
the ARGO-YBJ excess map is fitted with a two-dimensional Gaussian function in a square region of $10^{\circ}\times 10^{\circ}$, finding a source extension
$\sigma_{ext}$= 1$^{\circ}$.8$\pm$0$^{\circ}$.5.
The observed differential energy spectrum is $dN/dE =(2.5\pm0.4) \times 10^{-11}(E/1 TeV)^{-2.6\pm0.3}$ photons cm$^{-2}$ s$^{-1}$ TeV$^{-1}$, in the energy range 0.2-10 TeV.
The angular extension is consistent with that of the Cygnus Cocoon as measured by $Fermi$-LAT, and the spectrum also shows a good connection with the one measured in the 1-100 GeV energy range.
These features suggest to identify ARGO J2031+4157 as the counterpart of the Cygnus Cocoon at TeV energies.
The Cygnus Cocoon, located in the star-forming region of Cygnus X, is interpreted as a cocoon of freshly accelerated cosmic rays related to the Cygnus superbubble.
The spectral similarity  with Supernova Remnants indicates  that the particle acceleration inside a superbubble is similar to that in a SNR.
The spectral measurements from 1 GeV to 10 TeV allows for the first time to determine the possible spectrum slope of the underlying particle distribution. A hadronic model is adopted to explain the spectral energy distribution.

\end{abstract}


\section{Introduction}
The Cygnus region of the Galactic plane stands out in the Northern sky for the complex features observed in radio, infrared, X-rays and gamma-rays.
It contains a high density interstellar medium and is rich in potential cosmic ray acceleration sites, e.g., Wolf-Rayet stars, OB associations, and supernova remnants (SNRs).
This region is home of a number of GeV gamma-ray sources as revealed by $Fermi$-LAT \citep{nolan12}.
Several noteworthy TeV gamma-ray sources were also detected within the Cygnus region in the past decade.
A review of these sources can be found in our previous paper \citep{barto12a}.
Gamma-rays are tracers of non-thermal particle acceleration and propagation.
The Cygnus region is therefore a natural laboratory for studying the origin of cosmic rays, which remains a century-long puzzle.

Cygnus X in the constellation of Cygnus is the largest star-forming region in the Solar neighborhood.
It contains many distinct H II regions, a number of Wolf-Rayet stars and several OB associations, namely OB2, OB1 and OB9.
Most objects seen in this region are located at the distance of $\sim$1.4 kpc \citep{hanson03}.
In the direction of Cyg OB2, the  gamma-ray source TeV J2032+4130 was discovered by the HEGRA collaboration \citep{aharon02,aharon05} and later observed by the Whipple \citep{konop07}, MAGIC \citep{albert08} and VERITAS \citep{aliu14} imaging air shower Cherenkov telescopes (IACTs).
Its extension is estimated to be about 0$^\circ$.1 and the integral flux above 1 TeV is 3\%$-$6\% that of the Crab flux.
TeV J2032+4130 is positionally coincident with the GeV pulsar PSR J2032+4127 and could be the corresponding pulsar wind nebula (PWN) \citep{camilo09,aliu14}.
The sources MGRO J2031+41 and ARGO J2031+4157, detected by Milagro  and ARGO-YBJ extensive air shower (EAS) arrays around 20 TeV and 1 TeV, respectively, overlap with TeV J2032+4130, but the fluxes are about 10 times higher \citep{abdo07b,barto12a,barto13a}.

Using the $Fermi$-LAT data, \cite{acke11} found a highly extended gamma-ray source above 1 GeV, overlapping the TeV emission.
Assuming a symmetrical two-dimensional Gaussian shape, the source angular extension was estimated to be 2$^{\circ}.0\pm0^{\circ}.2$.
The emission was interpreted as a cocoon of freshly accelerated cosmic rays in the Cygnus superbubble, which was taken as  evidence for the long-advocated hypothesis that OB associations host cosmic ray factories.
The extended emission region will be denoted as Cygnus Cocoon in the following text.

The extension of MGRO J2031+41, obtained by Milagro with a later analysis, is 1$^{\circ}.8$,  similar to that of the Cygnus Cocoon. For such a complex region, \cite{abdo12} pointed out that the morphology of MGRO J2031+41  may be produced by the superposition of the central source TeV J2032+4130 and a more extended emission, which causes the large discrepancy between the fluxes measured by  IACTs and EAS  arrays. However, the correlation between the TeV extended emission and the Cygnus Cocoon remained unclear.

VERITAS surveyed the Cygnus region.  Due to the narrow field of view (FOV, typical 3$^{\circ}$.5), VERITAS could not detect the gamma-ray emission on the angular scale of the Cygnus Cocoon. In Cherenkov telescopes the background under a point-like source is estimated by the density of events in rings around the source. For an extended source this method makes the source to be self-subtracted  \citep{aliu13}.
At one corner of the Cygnus Cocoon,  VERITAS detected a new TeV gamma-ray source VER J2019+407 \citep{aliu13} in the direction of SNR G78.2+2.1 with an angular extension of 0$^{\circ}.23 \pm 0^{\circ}.05$ and a flux above 320 GeV of 3.7\% Crab units.
Associated to VER J2019+407, a marginal signal, denoted as ARGO J2021+4038, was reported by the ARGO-YBJ collaboration when surveying the northern sky \citep{barto13a}. This source should also partly contribute to the extended source MGRO J2031+41.

The ARGO-YBJ experiment is an EAS array  with a wide FOV, able to monitor the sky in the declination band from -10$^{\circ}$ to 70$^{\circ}$ \citep{barto13a}. Due to the full coverage configuration and the location at high altitude (4300 m a.s.l.), the detector energy threshold is $\sim$300 GeV, much lower than that of Milagro.
ARGO-YBJ can extend the measurements of the space-based experiment $Fermi$-LAT at 0.1$-$300 GeV to higher energies without any gap.
In this paper we reanalyze the ARGO-YBJ data recorded in the region of ARGO J2031+4157, to study the relation of this source with the Cygnus Cocoon detected by $Fermi$-LAT.

\section{The ARGO-YBJ experiment}
The ARGO-YBJ experiment, located at the Yangbajing Cosmic Ray Laboratory (Tibet, China, 90$^{\circ}$.5 East, 30$^{\circ}$.1 North), is designed for very high energy (VHE) gamma-ray astronomy \citep{aielli09a,aielli09b,aielli10,barto11,barto12a,barto12b,barto12c,barto13b} and cosmic ray \citep{aielli09e,barto12d,barto12e,barto13c} observations.
The detector, extensively described in \citep{aielli06,aielli09c}, consists of a $\sim$74$\times$ 78 m$^2$ carpet made of a single layer of Resistive Plate Chambers (RPCs) with $\sim$93$\%$ of active area, surrounded by a partially instrumented ($\sim$20$\%$) ``guard ring'' area up to $\sim$100$\times$110 m$^2$.
Each RPC (2.8 m $\times$ 1.25 m) is equipped with 10 logical pixels (called ``pads", 55.6 cm $\times$ 61.8 cm) used for triggering and timing purposes.
The arrival times of the particles are measured by 18,360 time-to-digital converters (TDCs) with a resolution of about
1.8 ns \citep{aielli09c}, which are calibrated using a software method  with precision of 0.4 ns \citep{he07,aielli09d}.

The  ARGO-YBJ experiment in its final configuration started taking data in  November 2007 and stopped in January 2013.
The trigger rate is 3.5 kHz with a dead time of 4\% and an average duty-cycle higher than $86\%$.
All the ARGO-YBJ data collected in that period are used in this analysis, for a total observation time of 1670  days.
Details of the detector performance and data analysis are widely discussed in \cite{barto13a}, where, using the same data sample, a survey of the Northern sky is presented, yielding 6 sources and 5 candidates, with an integrated sensitivity of 24\%  Crab units. The detector angular resolution for gamma rays ranges from $1^{\circ}.7$  to $0^{\circ}.2$, depending on the number of hit pads N$_{pad}$ \citep{barto11b,barto13a}. The sky map in celestial
coordinates (right ascension and declination) is divided into a
grid of $0^{\circ}.1\times0.1^{\circ}$ bins and filled with detected events. The   ``direct integral method" \citep{fley04} is adopted  to estimate the number of cosmic ray background events in each bin.  A smoothing is applied with an energy-dependent point spread function  \citep{barto13a}. The  Li$-$Ma method \citep{li83} is used to estimate the significance of the excess.

\section{Results}
The significance map around ARGO J2031+4157 as observed by ARGO-YBJ using the events with N$_{pad}\geq$20 is shown in Figure 1.
The highest statistical significance is 6.1 standard deviations (s.d.), corresponding to the position of ARGO J2031+4157.
The positions of other known TeV sources and of the Cygnus Cocoon are also marked in the figure.
Our excess almost fully fills the extended region of the Cygnus Cocoon, indicating their similar angular size.
At the top right corner of the excess, there is a 4.3 s.d. peak corresponding to ARGO J2021+4038, which is associated to VER J2019+407.

To derive the possible emission from the   Cygnus Cocoon, the contribution from the overlapping sources TeV J2032+4130 and VER J2019+407, and from the nearby sources VER J2016+371 \citep{aliu11} and MGRO J2019+37 \citep{abdo07b}  must be subtracted.
For these sources, the spectra and the angular extensions determined in \citep{aharon05,aliu13,aliu11} are used.
For MGRO J2019+37, the flux measured by Milagro is higher than the upper limits derived by ARGO-YBJ, although still within one standard deviation error.
Recently, MGRO J2019+37 was resolved into two VERITAS sources, namely VER J2016+371 and VER J2019+368.
So the spectra determined by VERITAS \citep{aliu11}, which have   better precision and  are consistent with both Milagro and ARGO-YBJ measurements, are used here.
We track the four sources path inside the ARGO-YBJ FOV and simulate the detector response in the gamma-ray energy range from 10 GeV to 100 TeV.
The detailed simulation of the ARGO-YBJ detector response to gamma-rays is realized by means of a code based on the GEANT package  \citep{guo10}.  The four sources contributions are removed before estimating the extension and spectrum of the Cygnus Cocoon.

In our previous analysis \citep{barto12a}, the angular extension of ARGO J2031+4157 was estimated by fitting the angular distribution of the events centered on MGRO J2031+41 within a radius of 2$^{\circ}$.2.
The excess events outside this region were considered as due to the Galactic diffuse gamma-ray emission.
Now, after the $Fermi$-LAT result indicating the presence of a large extended source, a larger region is used to evaluate the extension of ARGO J2031+4157.
To achieve a better angular resolution, only events with N$_{pad}\geq$60 are used.
Assuming for the source shape   a symmetrical two-dimensional Gaussian function, we fit the ARGO-YBJ excess in a square region of $10^{\circ}\times 10^{\circ}$  around  ARGO J2031+4157.
The result of the fit gives a source position with RA=(307.8$\pm$0.8)$^{\circ}$ and Dec=(42.5$\pm$0.6)$^{\circ}$, and an extension
$\sigma_{ext}$= (1.8$\pm$0.5)$^{\circ}$,
consistent with the angular size of the cocoon as measured by $Fermi$-LAT (2.0$\pm$0.2)$^{\circ}$,  within the statistical uncertainties (see Fig.1).
The dependence of this result on the source spectral energy distribution (SED)  is found to be negligible.

To study  the spectral behaviour of ARGO J2031+4157, the extension  $\sigma_{ext}=2^{\circ}$ and the position of Cygnus Cocoon determined by $Fermi$-LAT at 1$-$100 GeV \citep{acke11} are used.
The fitting method described in \cite{barto11} is adopted.
In this procedure, the path of the Cygnus Cocoon inside the ARGO-YBJ FOV is tracked during the ARGO-YBJ life time.
The expected emission is generated by sampling gamma-rays in the energy range 10 GeV - 100 TeV assuming a power-law function.
The variable used to determine the event energy is the number of hit pads N$_{pad}$.
The energy spectrum is estimated by comparing the detected signal and the expected signal in six N$_{pad}$ intervals: 20$-$39, 40$-$59, 60$-$99, 100$-$199, 200$-$499 and $\geq$500.
Before fitting, the contribution of the four nearby sources is removed.
According to our simulations, this contribution is dominated by the two sources TeV J2032+4130 and VER J2019+407,
and is equal to 13.2\%, 11.1\%, 12.1\%, 10.4\%, and 16.2\%, respectively, in the first five  N$_{pad}$ intervals (in the 6$^{th}$ interval there is no excess).

The best fit to the spectrum ($\chi^{2}$/dof = 2.4/4) is:
 \begin{equation}
\frac{dN}{dE}=(2.5\pm0.4) \times 10^{-11}(E/1 TeV)^{-2.6\pm0.3} \ \ photons \ cm^{-2} s^{-1} \ TeV^{-1}.
\end{equation}
The integral flux above 1 TeV is (1.52$\pm$0.37)$\times 10^{-11}$ photons cm$^{-2}$ s$^{-1}$, corresponding to 0.8$\pm$0.2 Crab units.
The median energies of the six N$_{pad}$ intervals are 0.40, 0.64, 0.92, 1.4, 2.7, and 6.5 TeV, respectively.
The found spectrum and the corresponding 1 sigma errors   are shown in Figure 2. The highest energy point is a 95\% confidence level (c.l.) flux upper limit.
The flux is higher than in our previous report \citep{barto12a}, since a larger source region is considered here. This also gives us a hint that the extension of the source is really larger than our previous estimation.
The given errors on the flux are statistical. The systematic errors are estimated to be less than 30\% \citep{barto11}.

Note that to subtract the contributions of  TeV J2032+4130 and VER J2019+407, the gamma-ray fluxes  determined by IACTs are used.
Some unclear systematic discrepancy between EAS arrays and IACTs has been found when determining the flux of extended sources. It is worth noting that these two techniques have achieved a good agreement for  the point source Crab Nebula \citep{abdo12b,barto13a}.
The fluxes of MGRO J1908+06 and HESS J1841-055 measured by the EAS arrays  Milagro and ARGO-YBJ are about 2$-$3 times higher than that determined by IACTs \citep{barto12c,barto13b}.
Therefore we can not exclude the possibility that the fluxes of TeV J2032+4130 and VER J2019+407 are underestimated by IACTs.
In this case the flux of the Cygnus Cocoon determined here would be overestimated by about 20$-$30\%.
However, the angular sizes of TeV J2032+4130 and VER J2019+407 are smaller than those of MGRO J1908+06 and HESS J1841$-$055, hence the expected discrepancy should be smaller. In particular, for MGRO J2019+37/VER J2019+368, if we use the result of Milagro instead of the VERITAS one, the cocoon flux and extension change by less than 10\%.

Figure 3 shows all the spectral measurements by $Fermi$-LAT, ARGO-YBJ and Milagro. The Milagro data refer to the source MGRO J2031+41 \citep{abdo07,abdo07b,abdo09}, which should contain the contributions from the overlapping and nearby sources.
In  \citep{acke11} the flux of MGRO J2031+41 is corrected by subtracting the extrapolation of TeV J2032+4130 at 12 TeV.
This ``corrected'' value is also shown in Figure 3. We should also remind that the Milagro flux at 12 and 20 TeV was determined in a region of $3^{\circ}\times 3^{\circ}$, which is too small compared to the Cygnus Cocoon extension and could contain less than 40\% of the gamma-ray emission. For these reasons the flux of MGRO J2031+41 is reported in Figure 3 but is not used in the following discussion.

The flux determined by ARGO-YBJ appears consistent with the extrapolation of the $Fermi$-LAT spectrum.
Given the consistency of spectra and angular sizes, the major emission of ARGO J2031+4157 can be identified as the counterpart of the Cygnus Cocoon at TeV energies.
It is worth  noting that, given the ARGO-YBJ angular resolution, a detailed comparison with the morphology found by $Fermi$-LAT is meaningless.

The combined spectrum of $Fermi$-LAT and ARGO-YBJ is fitted ($\chi^{2}$/dof = 2.7/9) by the power law function
$dN/dE =(3.5\pm0.3) \times 10^{-9}(E/0.1 TeV)^{-2.16\pm0.04}$ photons cm$^{-2}$ s$^{-1}$ TeV$^{-1}$, as shown by the dot-dashed line in Figure 3.
The upper limits of Fermi-LAT and ARGO-YBJ indicate weak evidence for the presence
of a slope change or cutoff below $\sim$1 GeV and above $\sim$10 TeV, respectively.

\section{Discussion}
The angular size of about 2$^{\circ}$ places the Cygnus Cocoon among the most extended VHE gamma-ray sources.
At a distance of 1.4 kpc, the observed angular extension corresponds to more than 50 pc, making the Cygnus Cocoon the largest identified Galactic TeV source.
Such a large region can be related to different scenarios.
PWNs and SNRs contribute to most of the extended Galactic TeV sources identified up to now.
Towards the Cygnus Cocoon, two pulsars (PSR J2021+4026 and PSR J2032+4127) and one SNR (SNR G78.2+2.1) have been detected.
As remarked in \citep{acke11}, the PWNs powered by these two pulsars are unlikely to explain the cocoon emission, and SNR G78.2+2.1 could be too young to be the unique accelerator in the cocoon able to diffuse over the whole region.
However, PSR J2032+4127 and the Cygnus Cocoon are well coincident apparently, and
we cannot rule out the possibility that the cocoon emission  is from the yet-undiscovered remnant of a supernova that originated the pulsar.
The favored scenario in \citep{acke11} is the injection of cosmic rays via acceleration from the collective action of multiple shocks from supernovae and winds of massive stars, which form the Cygnus superbubble. Such  superbubbles have been long advocated as cosmic ray factories \citep{bykov01,pari04,ferr10}. Possibly, the Cygnus Cocoon is the first evidence supporting   such a hypothesis.

For such a large extended region, no significant morphology and spectrum variation have been found by \cite{acke11} in the range from 1 to 100 GeV.
The energy spectrum from 1 GeV to 10 TeV  shows a simple power-law shape, which is very similar to those of SNRs, such as Cassiopeia A, Tycho, W51C, IC 443, and so on \citep{yuan12}.
This indicates that the Cygnus Cocoon may be an unknown SNR, or that the particle acceleration inside a superbubble is similar to that in a SNR. No matter which accelerator is responsible for the Cygnus Cocoon emission, the whole spectral shape of the gamma-ray emission from 1 GeV to 10 TeV allows for the first time to determine a possible spectral slope of the underlying particle distribution. Different scenarios have been proposed to explain the emission mechanism of gamma-rays, which can be produced via inverse-Compton scattering of background photon fields by high energy electrons, or, in hadronic models, by $\pi_{0}$-decay from inelastic proton-proton or proton-photon interactions.
The electron bremsstrahlung can be ignored if the electron-to-proton ratio is about 1\% as measured around earth.
The close relation between the emission morphology and the interstellar structure revealed by \cite{acke11} favors a cosmic ray origin.
The $Fermi$-LAT measurement below 3 GeV is also a hint of the $\pi_{0}$-decay feature \cite{acke13}. Moreover, the gamma-ray spectrum predicted by IC process is always curved, and it is difficult for the pure leptonic model to produce such a simple power law shape from 1 GeV to 10 TeV.

In this work we adopt a purely hadronic emission model \citep{drury94} to produce the gamma-ray emission from the cocoon.
In our model, the observed gamma-rays are attributed  to the decay of $\pi_0$ mesons produced in inelastic collisions between accelerated protons and target gas. The predicted spectrum is shown as the thick solid line in Figure 3.
It is assumed that the primary proton spectrum follows a power law with index $\alpha$ and with
an exponential cutoff energy $E_{c}$, i.e., $E^{\alpha} e^{-E/E_{c}}$. The value of $\alpha$ is the same as the spectral index of gamma-rays and $E_{c}$ is set to 150 TeV. This $E_{c}$ value is the maximum allowed by the ARGO-YBJ upper limit.
It is worth recalling that, as explained in the last section, the data
points obtained by Milagro are not taken into account for this model.
Taking them into account, the  $E_{c}$  would be around 40 TeV, giving the
dotted line shown in Figure 3.
The total energy of the high-energy protons  is $1.5\times10^{50}\times(10/n)$ ergs  at a distance of 1.4 kpc, where $n$ is the gas density per cm$^3$ averaged over the emission volume.
The Cygnus region hosts a giant molecular-cloud complex with total mass of 8$\times10^6$ $M_{\odot}$ \citep{acke12}.
The interstellar gas density should be higher than 10 cm$^{-3}$. Hence, the derived proton power should be less than 1.5$\times10^{50}$ ergs, which can be reasonably provided by one supernova, which usually releases $\sim10^{51}$ ergs, about  10\% of which can be transferred to the accelerated particles. This result plausibly supports one SNR or several SNRs, i.e., a supperbubble, as the accelerator responsible for the Cygnus Cocoon emission.
The main feature of the gamma-ray emission from the hadronic model is that it predicts a  clear cutoff at energy below 1 GeV \cite{acke13}, which is significantly different from that from the leptonic model and can be checked by future long-term observation from $Fermi$-LAT.

In a cosmic ray source, both electron and nuclei can be accelerated at the same time. The gamma-ray spectrum of Cygnus Cocoon can also be reproduced by a hybrid model.  Recently, \cite{yuan12} proposed a unified model for the gamma-ray emission of SNRs.
In this model, they assumed a constant ratio (1\%) and identical spectral shape for cosmic-ray protons and electrons, which was derived from the locally observed spectra   by taking into account the transport effects from sources to the observers.
The diversity of gamma-ray spectra can be caused only by the differences of the medium gas density.
According to this model, when the differential spectral index of  gamma-rays is 2.16, the gamma-ray emission  below 100 GeV is dominated by pion decay and the emission above 100 GeV is dominated by the electron inverse Compton  scattering on low energy background photons.

\section{Conclusions}
Using the whole ARGO-YBJ data collected from 2007 November to 2013 January, we reanalyzed the TeV gamma-ray emission from   ARGO J2031+4157.
The source extension is found consistent with that of the Cygnus Cocoon determined by $Fermi$-LAT at 1$-$100 GeV.
The measured spectrum is consistent with a simple power-law extrapolation of the spectrum measured by $Fermi$-LAT, suggesting the same origin for both GeV and TeV gamma-ray emissions.
These features support the identification of ARGO J2031+4157 as the TeV counterpart of the Cygnus Cocoon.

Even if a yet-undiscovered nebula can not be ruled out, such a large extended emission is likely due to the collective action of multiple shocks in a superbubble.
The broad spectrum and the similarity with other SNRs indicate that the particle acceleration inside a superbubble is similar to that in a SNR.
The GeV-TeV spectrum allows for the first time to determine the possible spectral slope of the underlying particle distribution. To further investigate into the Cygnus Cocoon and its relation with TeV J2032+4130, more sensitive observations by detectors with large collection area, high angular resolution and wide field of view are needed. The new EAS experiments, such as HAWC, Tibet+MD, and LHAASO (more details about these detectors can be found in \citep{chen13}) are expected to play an important  role for such an issue.

\acknowledgments
 This work is supported in China by NSFC (No.10120130794, No.11205165, No.11375210),
the Chinese Ministry of Science and Technology, the
Chinese Academy of Sciences, the Key Laboratory of Particle
Astrophysics, CAS, and in Italy by the Istituto Nazionale di Fisica
Nucleare (INFN).

We also acknowledge the essential supports of W.Y. Chen, G. Yang,
X.F. Yuan, C.Y. Zhao, R. Assiro, B. Biondo, S. Bricola, F. Budano,
A. Corvaglia, B. D'Aquino, R. Esposito, A. Innocente, A. Mangano,
E. Pastori, C. Pinto, E. Reali, F. Taurino and A. Zerbini, in the
installation, debugging and maintenance of the detector.


\clearpage
\begin{figure}
\plotone{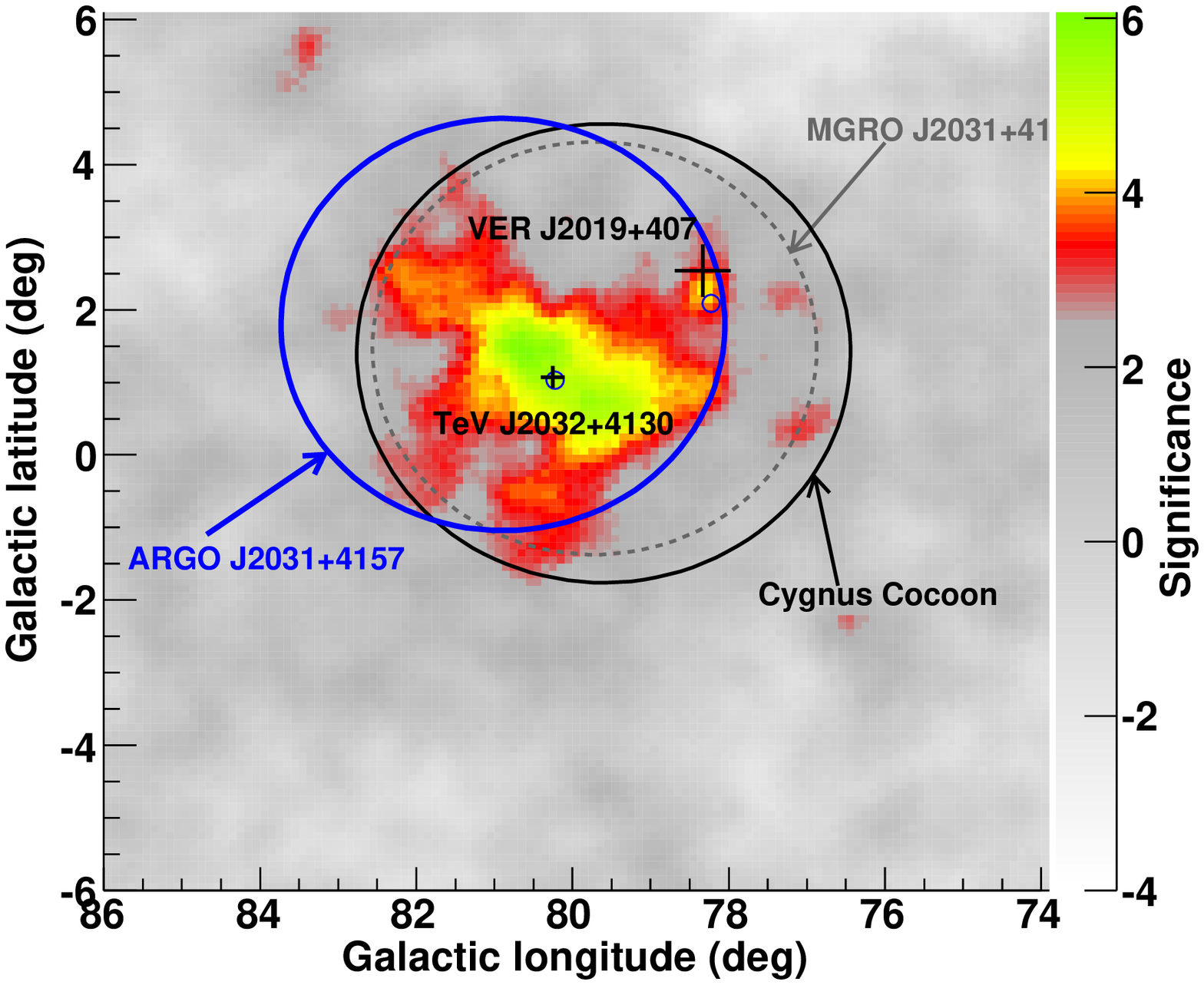}
\caption{
Significance map around ARGO J2031+4157 as observed by the ARGO-YBJ
experiment. The large circles indicate the positions of ARGO J2031+4157, MGRO J2031+41 and the Cygnus Cocoon, and the corresponding 68\%  containment regions \citep{acke11,abdo12}.
The position and  extension of TeV 2032+4130 and VER J2019+407  are  marked with crosses \citep{aharon05,aliu11,aliu13}.
The small circles indicate the positions of PSR 2021+4026 and PSR 2032+4127.
}
\label{fig1}
\end{figure}

\begin{figure}
\plotone{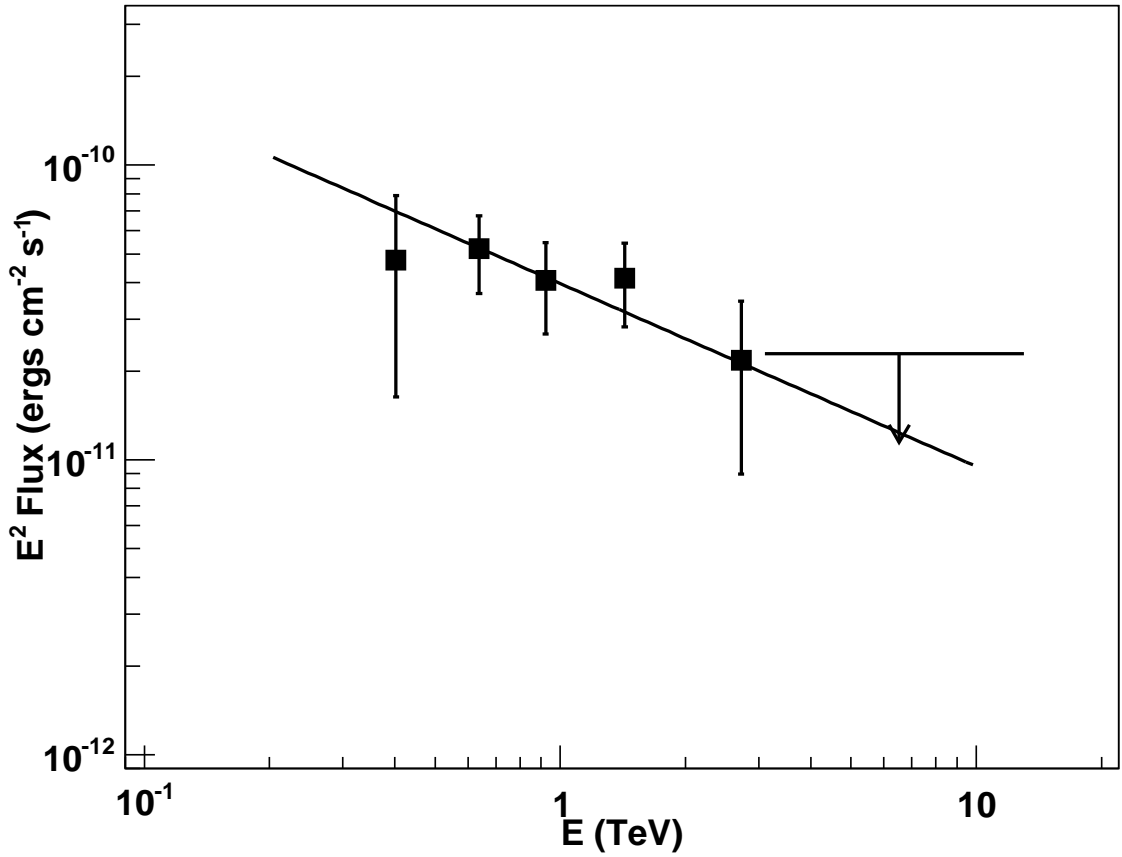}
\caption{Energy density spectrum of the Cygnus Cocoon as measured by the
ARGO-YBJ experiment. The
solid line shows the power-law fit to the data points. The arrow indicates the 95\% c.l. upper limit.
Only statistical errors are shown.}
\label{fig2}
\end{figure}

\begin{figure}
\plotone{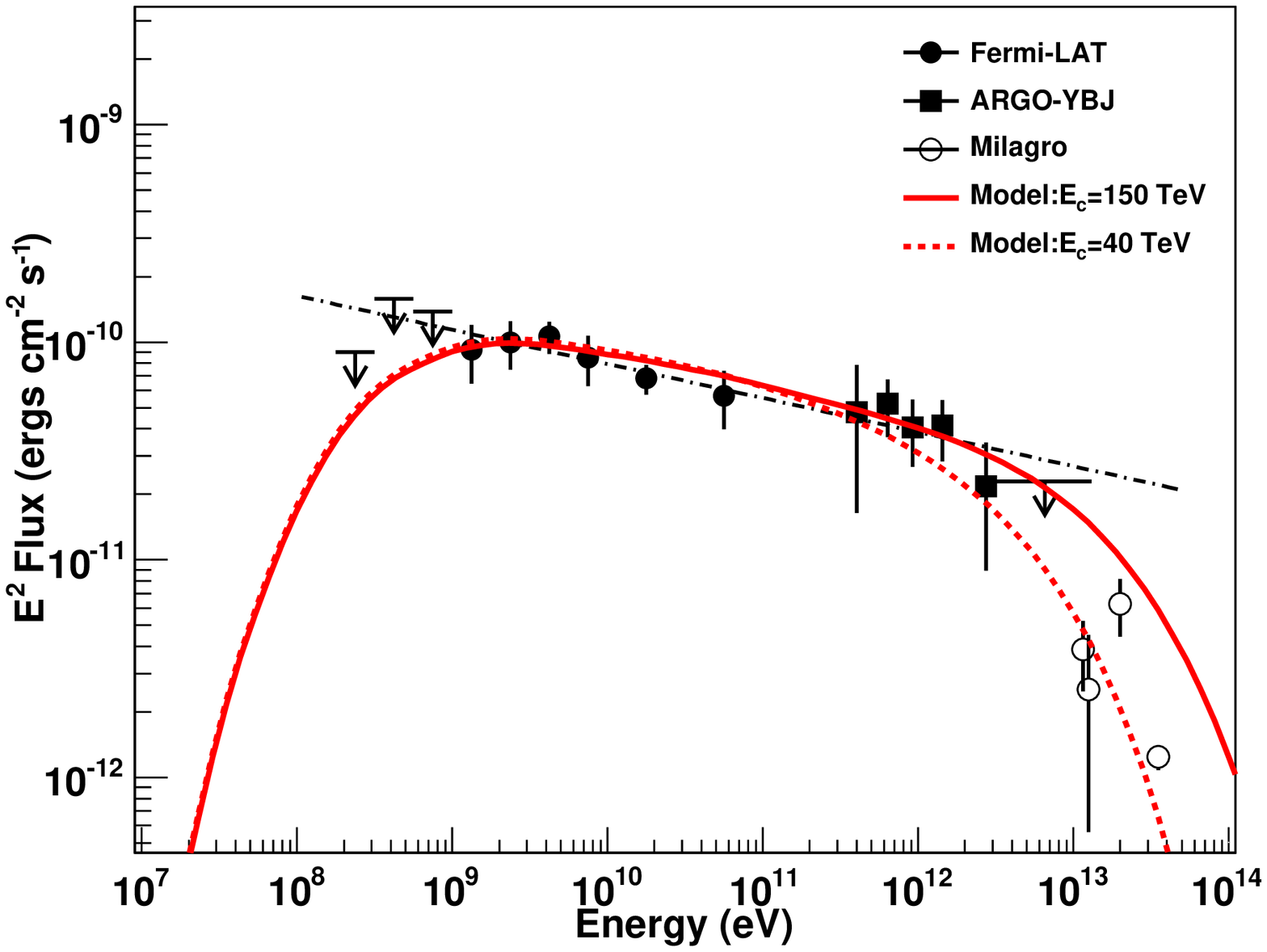}
\caption{ Spectral energy distribution of the Cygnus Cocoon by different detectors. The arrows below 1 GeV indicate the upper limits obtained by $Fermi$-LAT \citep{acke11}.
The   points at 12, 20 and 35 TeV are reported by Milagro for MGRO J2031+41 \citep{abdo07,abdo07b,abdo09}. The lower data point at 12 TeV is the Milagro flux after the subtraction of the TeV J2032+4130 contribution \citep{acke11}. The dot-dashed line shows the best fit to the $Fermi$-LAT and ARGO-YBJ data using a simple power-law function. The thick solid line is predicted by a hadronic model with a proton cutoff energy at 150 TeV. The  dotted line is predicted by a model with cutoff energy at 40 TeV.
}
\label{fig3}
\end{figure}

\clearpage

\end{document}